# Quantitative Measurement of adhesion energy between nanolayers and substrates using a nanowire-supported bridging method

Xiaodong Song[1], Lizhen Hou[1, *], Ruizhe Liu[1], Noman Akhtar[2], Peng Wang[1], Shiliang Wang[2]

1. School of Physics and Electronics, Hunan Normal University, Changsha, 410081, China
2. School of Physics, Central South University, Changsha, 410083, China

E-mail: lizhenhou@hunnu.edu.cn

## Abstract

The measurement of adhesion energy between nanolayers and substrates holds significant importance for the design, fabrication, and stability assessment of micro-/nanoscale devices relying on nanolayers. In this study, we propose a nanowire-supported bridging method based on an optical microscope-based nanomanipulation technique to quantitatively measure the adhesion energy between nanolayers and substrates. Using this innovative approach, we conducted adhesion energy measurements between mica nanolayers and Si substrates, revealing a value of approximately 110 J/m$^2$. Additionally, we discuss the applicable conditions of this new method. The proposed technique allows measurements in atmospheric conditions and is, in principle, applicable to all types of nanolayers and substrates. Consequently, it holds promise as a universal method for assessing adhesion energy between nanolayers and substrates, considering environmental factors such as atmosphere and roughness.

## 1. Introduction

Mica is a sheet-like silicate mineral with a unique crystal structure having the chemical formula of $KAl_2(Si_3Al)O_{10}(OH)_2$. It comprises a wide range of fascinating properties such as temperature stability, visible-light transparency, electric insulation, ultraviolet (UV)-shielding, atomic level flatness, and chemical durability [1-3]. Recently, several techniques are used for the fabrication of mica nanolayers (MNLs) and mica nanosheets, including mechanical exfoliation technique [4], sonication exfoliation [5], intercalation-promoted exfoliation technique [6], microwave-expanded exfoliation technique [7, 8]. Mica nanolayers (MNLs) have remarkable mechanical properties (High young's modulus, higher breaking force) and excellent electronic transport, flexibility, transparency, thermal and chemical stability, and high dielectric constant. As a result, MNLs are potential candidate for excellent mechanical applications, including flexible ultrathin insulating substrates/dielectrics or for reinforcement in nanocomposites and in device fabrication such as field-effect transistors, electrodes, and plasmonic detectors [9-14]. While the consistency of a flexible device is considerably affected by the detachment within MNLs or with other nanolayers due to the interfacial stresses prompted by mechanical bending, twisting and stretching, thermal loading, surface effect, and other environmental factors during device usage [15][16]. Therefore, understanding the contact behaviour of MNLs and other materials is essential for the rational use of mica-based devices for stable use. Due to the relationship between film thickness (number of layers) and stiffness, the boundary between membrane and plate becomes blurred as the film thickness decreases [17, 18]. Therefore there is a distinction between film theory and plate theory in the study of adhesion energy. Whether there is a clear boundary and a parameter to express it is a question that many people want to solve.

To address and eliminate all these concerns and for an enhanced commercial use of flexible electronic devices, a complete investigation of the adhesion of nanofilms/nanolayers is most inevitable, while considering the adhesion for flat surfaces it is still challenging. The study of the adhesion phenomenon could be traced back to at least Bowden and Bastow's work on surface forces in the 1930s [19]. So far, several methods for measuring the interfacial adhesion energy have been developed, including the nano-scratch technique based on atomic force microscopy (AFM) [20, 21], in situ nanomanipulation techniques[22]，wrinkle-based method [23, 24], blister test [25-27]and bridging method[28]. In the previous study, The nano-scratch technique has the problem of the limitation of the tip material and the uncertainty of the contact area between the tip and the sample, the mechanical in-situ stripping method has to put the sample under high-energy electron beams, which can not avoid the effect of the electron beams on the interfacial adhesion, the blister method can not rule out the effect of the air inside the chamber on the bubble profile. To overcome all these challenges, in our previous work we presented a new buckle(nanowire-supported ) method for the measurement of the adhesion of MNLs, based on the optical microscopic (OM) nanomanipulation under controlled environmental conditions of different values of relative humidity and temperature. The nanomanipulation-based bridging method uses the shape of the buckle to correlate with the adhesion energy [29],but the stored bending energy in the bridging method reaches the equilibrium of the maximum static friction of the stored bending energy did not reach the result of the calculation of the results of a small. We have reconstructed the new test system using nanowires as supports, and now we have further reduced the thickness of the film to below 100nm. As the film thickness decreases, for the theoretical part of the calculation of the adhesion energy a film stretching energy term was added to accommodate the thinner mica of this experiment in addition to the usual consideration of the bending energy stored by film bending only. We also explored the transition from bending plates to stretching films for the extent to which the bending and stretching energies contribute to each other.

## 2. Methodology

### 2.1 Formation of nanowire-supported MNL nanobridge

The preparation of MNLs involves the following peeling method: Start by using Japanese Nitto NITTO (448S) tape, cutting 7-8 sections, each measuring 10 centimeters, for subsequent use. Then, adhere the tape to the surface of the mica substrate and peel it off from the substrate, leaving the tape with a layer of mica, exhibiting bright colored light reflection. Next, take the remaining tape, affix it to the initial tape with adhered mica layer, and press firmly to ensure complete coverage of the mica layer. Quickly peel off the tape with both hands, repeating this process to obtain a gradually thinner mica layer, as shown in Figure 1(a). As the mechanical peeling progresses, the colored light diminishes. Continue peeling until the mica layer on the tape no longer reflects colored light.

The preparation of nanowire-supported MNL bridges by using the optical microscopy (OM) nanomanipulation technique : A Si wafer (4×4 mm) was ultrasonically cleaned using acetone, alcohol, and deionized water in turn, and then the mica layers mechanically peeled in (1) are attached to the Si substrate and left to stand for a few minutes in Figure 1(b) and 1(c). By using an electrically chemical etched W needle tip mounted onto a 3D nanopositioner(P-616.3C, Physik Instrumente; closed-loop resolution: ~ 0.4 $nm$; travel range: 100 $\mu m$/axis; linearity error: 0.03%), a selected MNL is picked up and placed above a SiC nanowire that was supported by the Si substrate, and then covered over the nanowire by detaching the W tip to form the nanowire-supported bridge as shown in Figure 1(d) - 1(f). Note that the entire manipulation process was performed under the OM (Objective lens: Mitutoyo M Plan APO 50× with a resolution of 0.4 μm), the formation of nanowire-supported MNL bridge could be monitored and recorded in real-time.

### 2.2 Mechanical model for calculating adhesion energy

Figure 2 schematically shows the cross-sectional view of a Si nanowire supported MNL nanobridge over a smooth substrate. Simplifying the MNL-substrate system as a 2D model [28], the profile of the MNL nanobridge can be determined by the balance between the interfacial adhesion and the bending of the MNL, i.e., the sum of the bending energy ($U_b$) in the NML and the interfacial adhesion energy of the nanowire-substrate system ($U_a$) arrives the minimum value. Taking the horizontal direction as the *x*-axis, the origin of the *x*-axis is at the position of the nanowire, and assuming the height and the half-span of the nanobridge are $w_0$ and $s$ (Figure 2), the deflection of the right side the nanobridge at equilibrium can be describe by [30],

$$w(x) = w_0 \left( \frac{3x^2}{s^2} - \frac{2x^3}{s^3} \right), \qquad (1)$$

The bending energy of the right side of the nanobridge can be written by [30, 31]

$$U_b = \frac{EI}{2} \int_0^s \left( \frac{\partial^2 w(x)}{\partial x^2} \right)^2 dx = \frac{Et^3 w_0^2 b}{2s^3}, \qquad (2)$$

where $E$ (= 190 GPa) is the Young's modulus of the MNL [32], $I$ and $b$ are the moment of inertia of and the width the MNL nanobridge, respectively. The interfacial adhesion energy in the contact area (right side) can be obtained by,

$$U_a = -\gamma b (l_0 - s), \qquad (3)$$

According to the principle minimum energy, we have, $\partial(U_a + U_b)/\partial s = 0$, should have the minimum value. As a result, the adhesion energy can be expressed as,

$$\gamma_b = \frac{3 E w_0^2 t^3}{2 s^4}. \qquad (4)$$

## 3. Results and discussion

Figure 3(a) shows the OM image of two MNLs covered over four SiC nanowires (NW1 to NW4) on a Si substrate, constructing four nanowire-supported nanobridges labelled as B1, B2, B3, and B4, respectively. Note that NW1 and NW3 were cut from the same long CVD-grown SiC nanowire, while NW2 and NW4 were cut from another long SiC nanowire. This suggests B1 and B3 should have the same or quite similar height, while nanobridge B2 and B4 should also have the same or quite similar height. Figure 3(b) and 3(c) display the corresponding AFM images of the four nanowire-supported MNL nanobridges. Cleary, the MNLs have very smooth surfaces, and the nanobridges have uniform profiles along the length directions of the beneath nanowires. To reveal their geometric structures, we plotted the 2D cross-sectional profiles of B1 to B4 based the AFM measurement, as shown in Figure 3(d) to 3(g), respectively. B1 and B2 were constructed using the same MNL, which has a thickness of $t_{12} \approx 194$ nm as shown in the insets of Figure 3(d) and 3(g). As the diameters of supporting nanowires, $d_{NW1} \approx 74$ nm and $d_{NW2} \approx 192$ nm, beneath the two nanobridges differ, the profiles and spans of the two bridges also differ, as shown in Figure 3(d) and 3(e). Similarly, B3 and B4 were constructed by the MNL with a thickness of $t_{34} \approx 74$ nm, and their bending profiles and spans also differ as the supporting nanowires have different diameters, $d_{NW3} = 75$ nm and $d_{NW4} = 193$ nm, as can be seen from Figure 3(f) and Figure 3(g). By fitting their cross-sectional profiles using Eq. (1), the heights and half-spans of B1 to B4 could be determined to be $w_{0-B1} = 74 \pm 2$ nm and $s_{B1} = 3075 \pm 46$ nm, $w_{0-B2} = 195 \pm 4$ nm and $s_{B2} = 4768 \pm 85$ nm, $w_{0-B3} = 76 \pm 1$ nm and $s_{B3} = 1556 \pm 10$ nm, $w_{0-B4} = 193 \pm 8$ nm and $s_{B4} = 2108 \pm 11$, respectively. Then, substituting the measured values of $w_0$, $t$ and $s$ into Eq. (4), the adhesion energies (per unit area) of $\gamma_{B1} = 0.127 \pm 0.011\ J/m^2$, $\gamma_{B2} = 0.153 \pm 0.017\ J/m^2$, $\gamma_{B3} = 0.114 \pm 9\ J/m^2$ and $\gamma_{B4} = 0.216 \pm 0.075\ J/m^2$ could be obtained for B1 to B4, respectively.

In theory, the van der Waals (vdW) adhesion energy for the MNL-Si system can be estimated using the following equation [33],

$$\gamma_{vdw-mica/Si} = \frac{A_{mica/Si}}{12\pi D^2}. \tag{5}$$

Here, $A_{mica/Si} = \sqrt{A_{mica} \times A_{Si}}$ represents the Hamaker constant for the interaction between mica and Si substrate, where $A_{Mica} = 13.5 \times 10^{-20}$ J and $A_{Si} = 18 \times 10^{-20}$ J are the Hamaker constants for mica and silicon, respectively [34]. The cut-off distance ($D$) is set to 0.2 nm in our calculations, considering both the MNL and Si have atomically flat surfaces in our test. Consequently, the vdW adhesion energy for our MNL-Si system could be determined to be $\gamma_{vdw-mica/Si} \approx 103$ mJ/m$^2$. Meanwhile, as the mechanically exfoliated MNLs exhibit discrete K$^+$ ions on their surface, this may induce electrostatic forces between the MNLs and the contacted Si substrate. Consequently, the actual adhesion energy could surpass the vdW adhesion energy, contingent on the specific testing conditions [35, 36]. For instance, an adhesion energy of 119.69 $\pm$ 20.47 mJ/m$^2$ was recorded between MNLs with a thickness of approximately 260 nm and Si in ambient air. Notably, in our experiments, only $\gamma_{B1}$ for B1 and $\gamma_{B3}$ for B3 closely align with the values reported in a previous study.

As B1 and B2 were constructed using the same MNL, it was anticipated that the measured adhesion energy values from them would be very close. However, the actual measurements reveal a deviation of over 20%. Notably, for the same MNL-substrate system, the measured adhesion energy values, $\gamma_{B3}$ from B3 and $\gamma_{B4}$ from B4, exhibit a substantial deviation of approximately 90%. This implies that the obtained adhesion energies are likely dependent on the specific profiles of the nanobridges. According to previous studies [37, 38], the presence of substantial values for the ratio of height to half-span, $w_0/s$, may result in an overestimation of the adhesion energy. This is attributed to the fact that Eq. (2) is obtained in the framework of linear elastic theory, whereas the elevated $w_0/s$ value could lead to the non-linear elastic deformation. Moreover, the tensile stress in the MNL may be associated with the thickness and specific profile of the corresponding nanobridge. In other words, the adhesion energy derived from Eq. (2) should be highly likely to be influenced by the ratio of $t/w_0$ or $t/s$. This underscores the importance of considering not only the uniformity of the MNL construction but also the specific geometric characteristics of the nanobridges when assessing adhesion energy values.

To reveal the dependence of the adhesion energy obtained from Eq. (2) on the specific geometric characteristics of the nanowire-supported nanobridges, we tested a total of 30 different nanobridges constructed on the same Si substrate and summarized the results in Figure 4. It is found that the adhesion energy may slightly increase from ~120 to ~150 mJ/m² as the thickness decreases from 472 to 42 nm, as depicted in Figure 4(a). This suggests that the thickness should not significantly affect the adhesion energy between the MNLs and Si substrates, which is also consistent with the previous results of 119.69 ± 20.47 mJ/m² obtained for MNLs with thickness above 260 nm [39]. Nevertheless, we also noted that the measured adhesion energy exhibits significant variability even for thicknesses that are quite similar. For instance, within the narrow thickness range of 42 to 76 nm, the adhesion energy shows a random distribution ranging from 0.10 to 0.22 J/m² (refer to the dashed rectangle region in Figure 4(a)). Meanwhile, it is interesting to found that adhesion energy in this narrow thickness range demonstrates a noticeable decreasing trend with the increase of the $t/w_0$ ratio (see the inset at the top right of Figure 4(a)). In fact, a closer examination of the $\gamma \sim w_0$ and $\gamma \sim t/w_0$ curves corresponding to the 30 nanobridges depicted in Figures 4(b) and 4(c), respectively, reveals that the adhesion energy remains unaffected by $w_0$. However, it does exhibit a decrease from ~ 210 to ~ 100 J/m² as the $t/w_0$ value increases from 0.35 to 3.50. In addition, it is also observed that the adhesion energy clearly increases from ~ 0.1 to ~ 0.2 J/m² as the height-to-half-span ratio ($w_0/s$) rises from ~ 0.02 to ~ 0.1, as shown in Figure 4(d).

Prior studies have proposed that the $\gamma$ value, calculated using Eq. (4), may be notably overestimated in two cases: (1) when the $w_0/s$ value exceeds 0.1, the linear elastic model is invalid; (2) when the bending stiffness of nanolayers is too small, the plate model becomes inapplicable. In our test, as the $w_0/s$ value measured is noticeably smaller than the well-established criterion of 0.1 for a significant contribution of nonlinear elastic deformation, it is suggested that the observed dependencies of γ on $t/w_0$ and $t/w_0$ should not be ascribed to nonlinear elastic deformation. It is believed that observed dependencies of γ on $t/w_0$ and $t/w_0$ should arise from the insufficient bending stiffness of nanolayers, leading to the failure of the plate theory; however, more detailed theoretical derivations await further rigorous confirmation. Nevertheless, our experimental results here suggest that the plate theory remains effective when the $t/w_0$ ratio is greater than 2, providing reasonably accurate measurement results.

## 4. Conclusion

In this study, we propose a novel nanowire-supported bridging method based on an optical microscope-based nanomanipulation technique for the quantitative measurement of adhesion energy between nanofilms/nanolayers and substrates. Utilizing this innovative approach, we measured the adhesion energy between mica nanolayers and Si substrates, revealing a value of approximately 110 J/m$^2$. Additionally, we observed that the plate theory for calculating film thickness becomes less reliable when the ratio of the measured film thickness to the nanowire diameter is around 2 or below. This discrepancy results in higher calculated values. However, when the ratio exceeds 2, the plate theory provides more accurate results. The proposed method is applicable for measurements in atmospheric conditions and, in principle, suitable for all types of nanofilms/nanolyaers and substrates. Therefore, it holds promise as a universal approach for assessing adhesion energy between nanofilms and substrates, considering environmental factors such as atmosphere and roughness.

**Acknowledgements**

This project is financially supported by the National Natural Science Foundation of China (No. 12072111), Hunan Provincial Natural Science Foundation of China (No. 2020JJ4676).
.

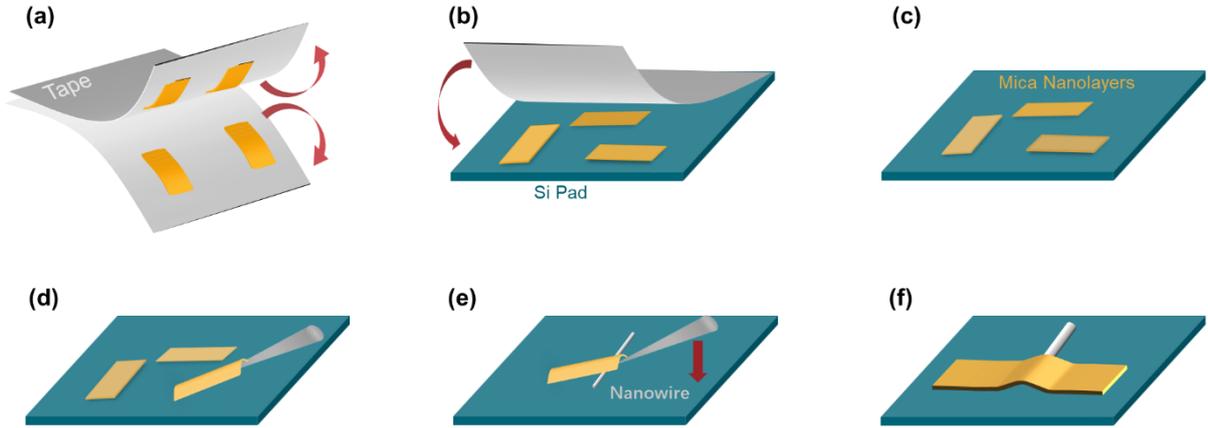

Figure 1. The process for preparing the nanowire-supported MNL bridges: (a) MNLs obtained by mechanical peeling; (b, c) MNLs attached onto the Si substrate; (d, e) pick-up, transfer and position of a MNL by a W needle tip; (f) formation of nanowire-supported MNL nanobridge.

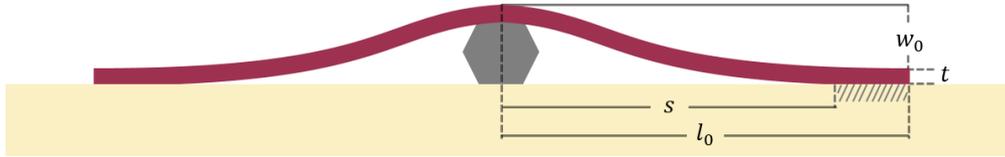

Figure 2. Cross-sectional profile of a nanowire-supported MNL nanobridge

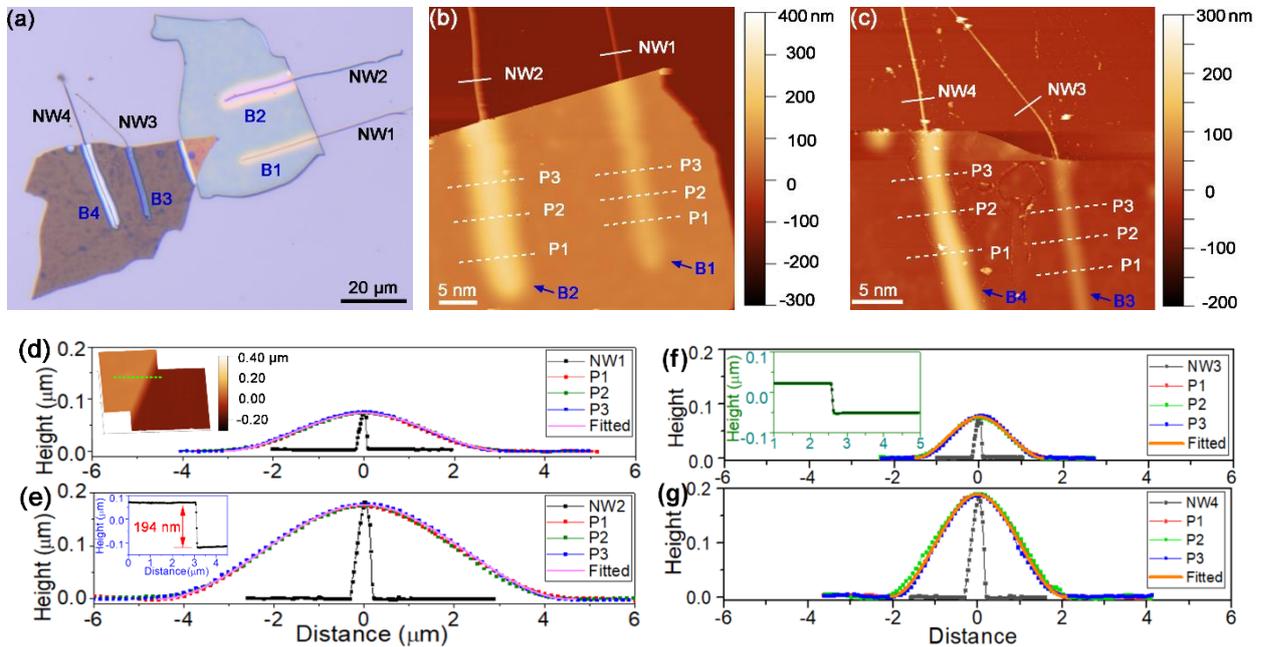

Figure 3. (a) OM image of four nanowire-supported MNL nanobridges labelled as B1 to B4, respectively. (b, c) AFM images of the B1 and B2, B3 and B4, respectively. (d, e) Cross-sectional profile of B1 and B2, respectively. The insets in (d) and (e) show the 3D and 2D profiles of the edge of the MNL in (b). (f, g) Cross-sectional profile of B3 and B4, respectively. The insets in (f) shows the 2D profiles of the edge of the MNL in (c).

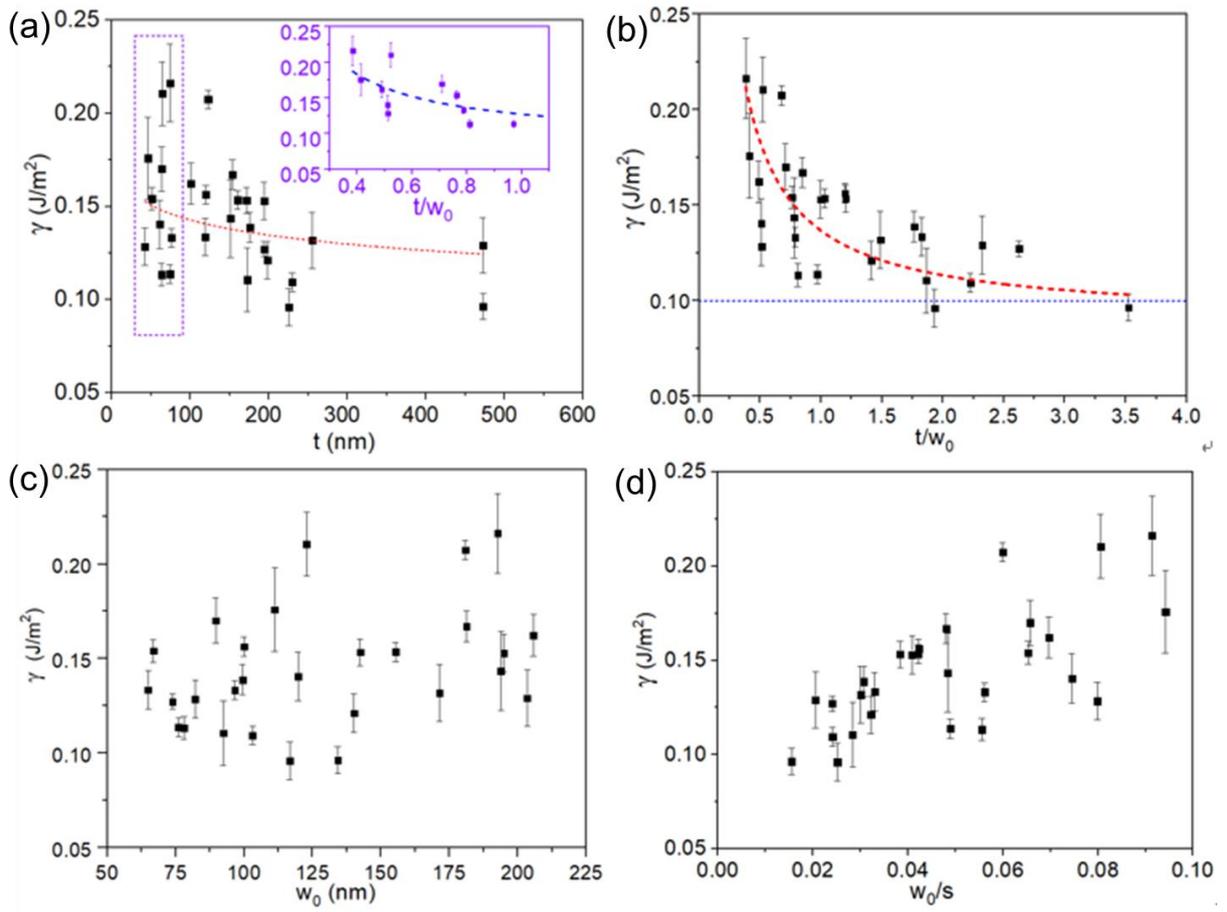

Figure 4. Dependence of adhesion energy on (a) the thickness, (b) the thickness to height ratio, (c) the height, and (d) the height to half-span of nanobridges.